\documentclass[%
longbibliography,
 reprint,
superscriptaddress,
 amsmath,amssymb,
 aps,
floatfix,
]{revtex4-1}
\usepackage[english]{babel}
\usepackage{comment}
\usepackage{natbib}
\usepackage{ulem}
\usepackage{graphicx}
\usepackage{dcolumn}
\usepackage{bm}
\usepackage{amsmath}
\usepackage{siunitx}
\usepackage{listings}
\usepackage{mathtools}
\usepackage{float}

\usepackage{xcolor}
\usepackage{url}
\usepackage{hyperref}
\hypersetup{
    colorlinks=true,allcolors=blue,}

\begin{document}
\preprint{APS/123-QED}

\title{Tuning Dzyaloshinskii-Moriya Interaction in Ferrimagnetic GdCo: A First Principles Approach}

\author{Md Golam Morshed}
\email{mm8by@virginia.edu}
\affiliation{Department of Electrical and Computer Engineering, University of Virginia, Charlottesville, VA 22904 USA}
\author{Khoong Hong Khoo}%
\affiliation{Institute of High Performance Computing, Agency for Science, Technology and Research, 1 Fusionopolis Way, Connexis, Singapore 138632, Singapore}
\author{Yassine Quessab}%
\affiliation{Center for Quantum Phenomena, Department of Physics, New York University, New York, NY 10003 USA}
\author{Jun-Wen Xu}%
\affiliation{Center for Quantum Phenomena, Department of Physics, New York University, New York, NY 10003 USA}
\author{Robert Laskowski}%
\affiliation{Institute of High Performance Computing, Agency for Science, Technology and Research, 1 Fusionopolis Way, Connexis, Singapore 138632, Singapore}
\author{Prasanna V. Balachandran}%
\affiliation{Department of Materials Science and Engineering, University of Virginia, Charlottesville, Virginia 22904 USA}
\affiliation{Department of Mechanical and Aerospace Engineering, University of Virginia, Charlottesville, Virginia 22904 USA}
\author{Andrew D. Kent}%
\affiliation{Center for Quantum Phenomena, Department of Physics, New York University, New York, NY 10003 USA}
\author{Avik W. Ghosh}%
\affiliation{Department of Electrical and Computer Engineering, University of Virginia, Charlottesville, VA 22904 USA}
\affiliation{Department of Physics, University of Virginia, Charlottesville, VA, 22904 USA}

\date{\today}

\begin{abstract}
We present a systematic analysis of our ability to tune chiral Dzyaloshinskii-Moriya Interactions (DMI) in compensated ferrimagnetic Pt/GdCo/$\text {Pt}_{1-x}\text {W}_x$ trilayers by cap layer composition. Using first principles calculations, we show that the DMI increases rapidly for only $\sim 10\%$ W and saturates thereafter, in agreement with experiments. The calculated DMI shows a spread in values around the experimental mean, depending on the atomic configuration of the cap layer interface. The saturation is attributed to the vanishing of spin orbit coupling energy at the cap layer and the simultaneous constancy at the bottom interface. Additionally, we predict the DMI in Pt/GdCo/X ($\text X=\text {Ta}, \text {W}, \text {Ir}$) and find that W in the cap layer favors a higher DMI than Ta and Ir that can be attributed to the difference in \textit{d}-band alignment around the Fermi level. Our results open up exciting combinatorial possibilities for controlling the DMI in ferrimagnets towards nucleating and manipulating ultrasmall high-speed skyrmions.
\end{abstract}
\pacs{Valid PACS appear here}
\maketitle
\begin{figure}[!htbp]
\includegraphics[width=\linewidth]{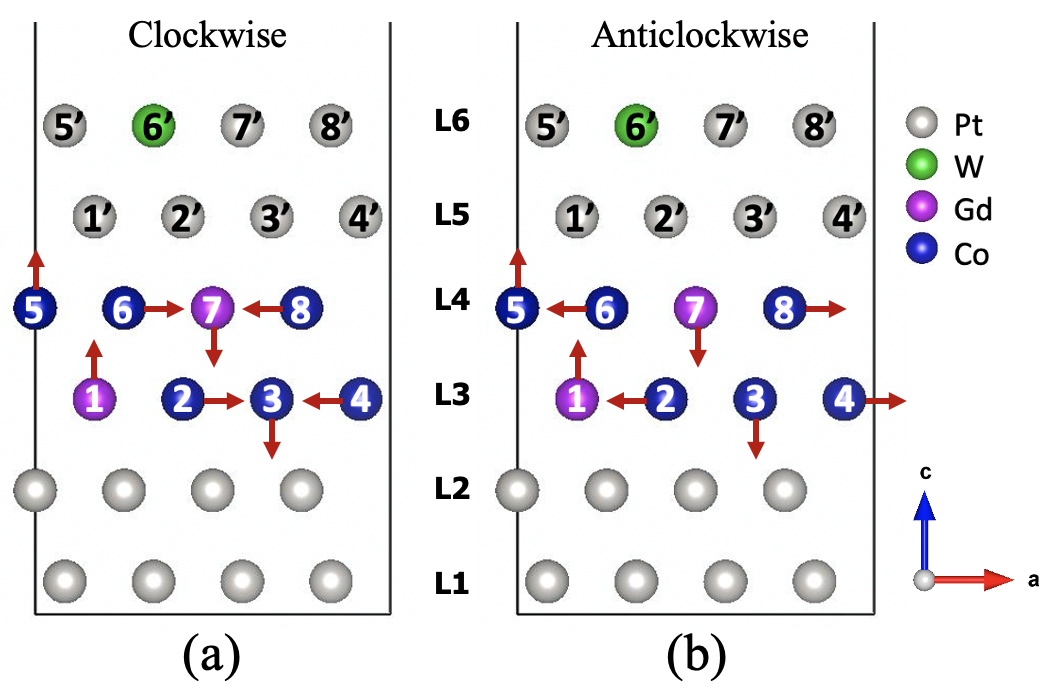}
\caption{Schematic of $\text{Pt}(2)/\text{Gd}\text{Co}(2)/\text{Pt}_{1-x}\text{W}_x(2)$ structure (number in parentheses denoting the number of monolayers) corresponding to $x=12.5\%$ for (a) CW, and (b) ACW spin configurations. The red arrows in the figure show the spin orientations. $\text{L1},...,\text{L6}$ denote the layer number while numbers in circles label atomic positions.}
\label{structure}
\end{figure}

\begin{figure*}[!htbp]
\includegraphics[width=.32\textwidth]{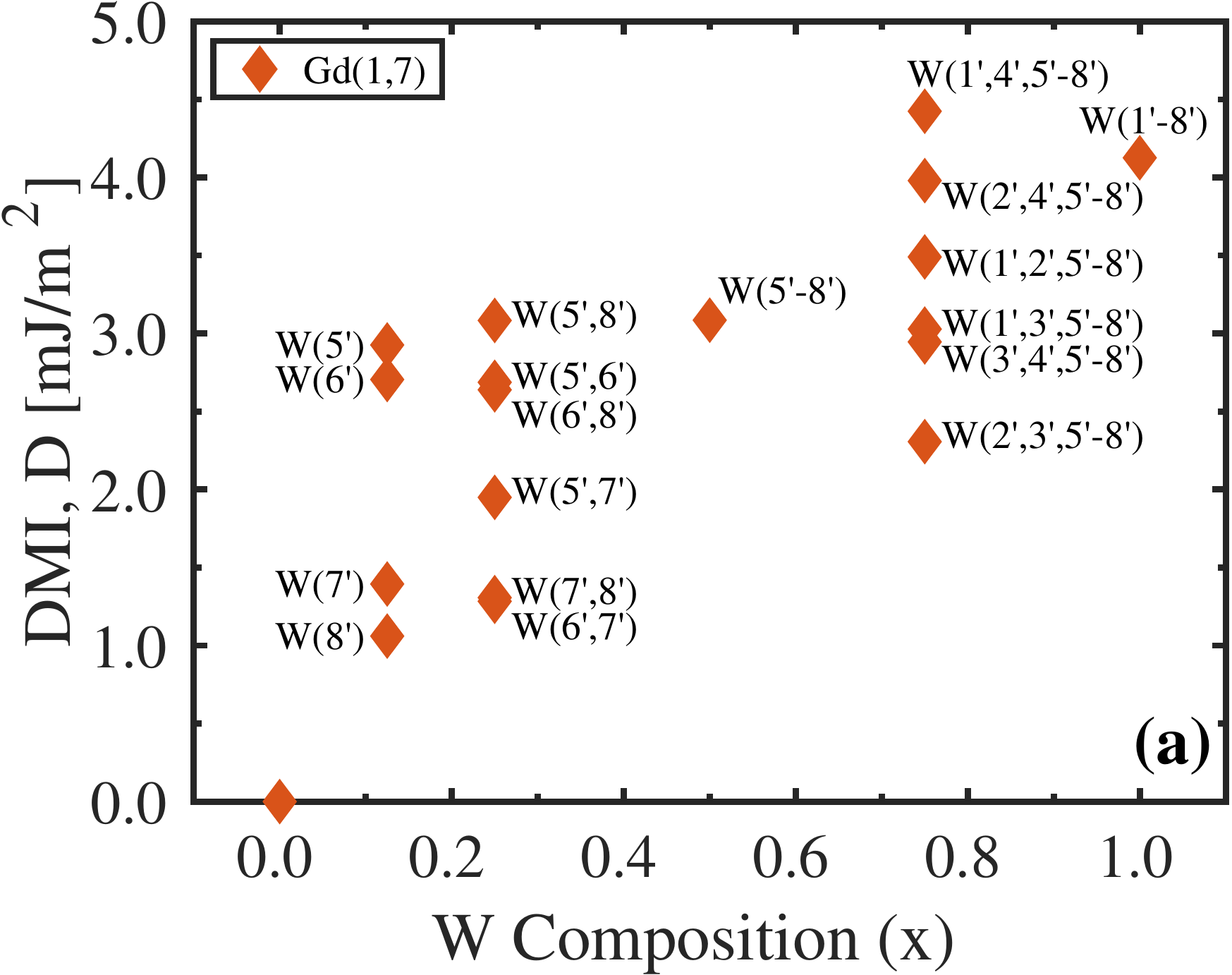}
\includegraphics[width=.32\textwidth]{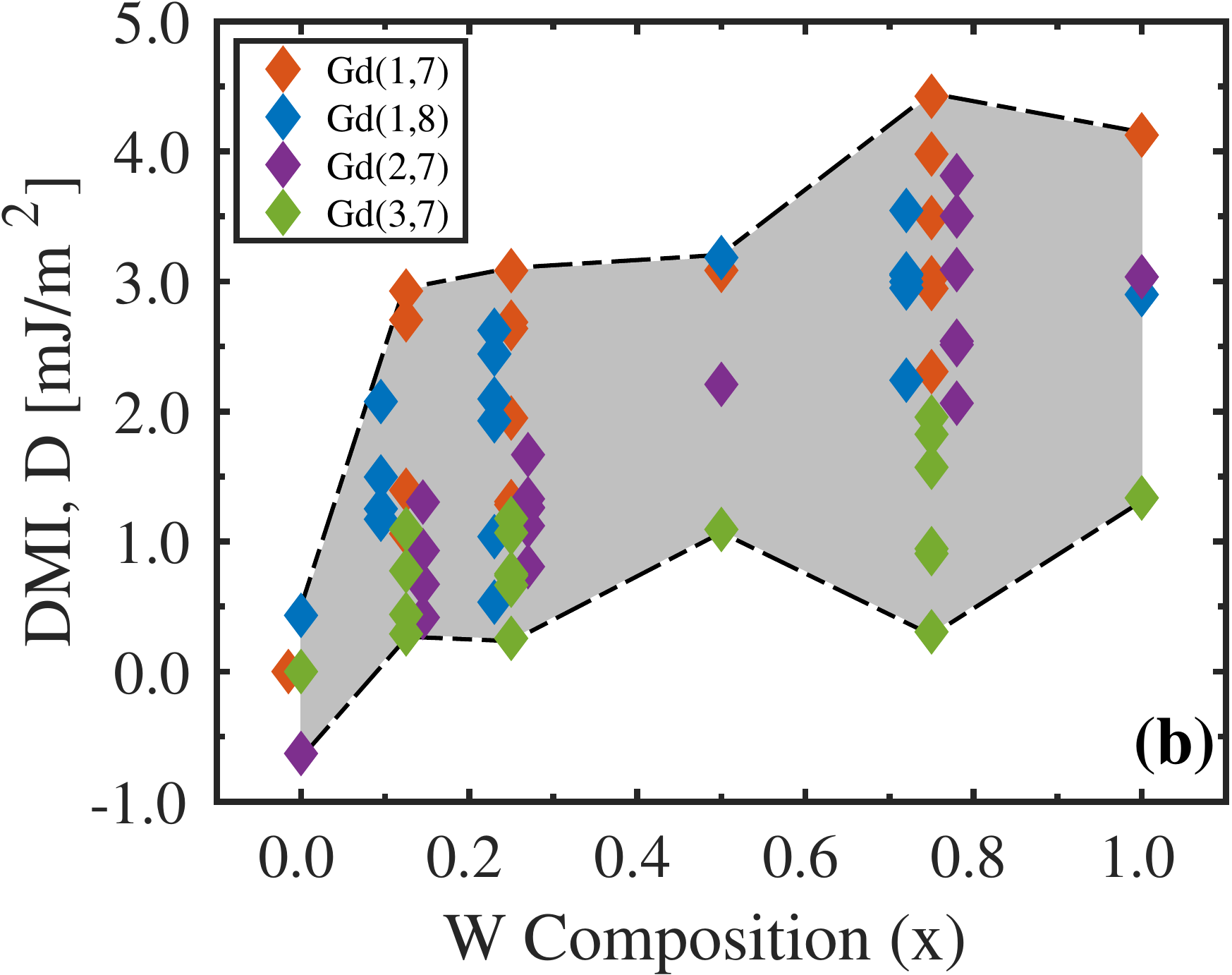}
\includegraphics[width=.32\textwidth]{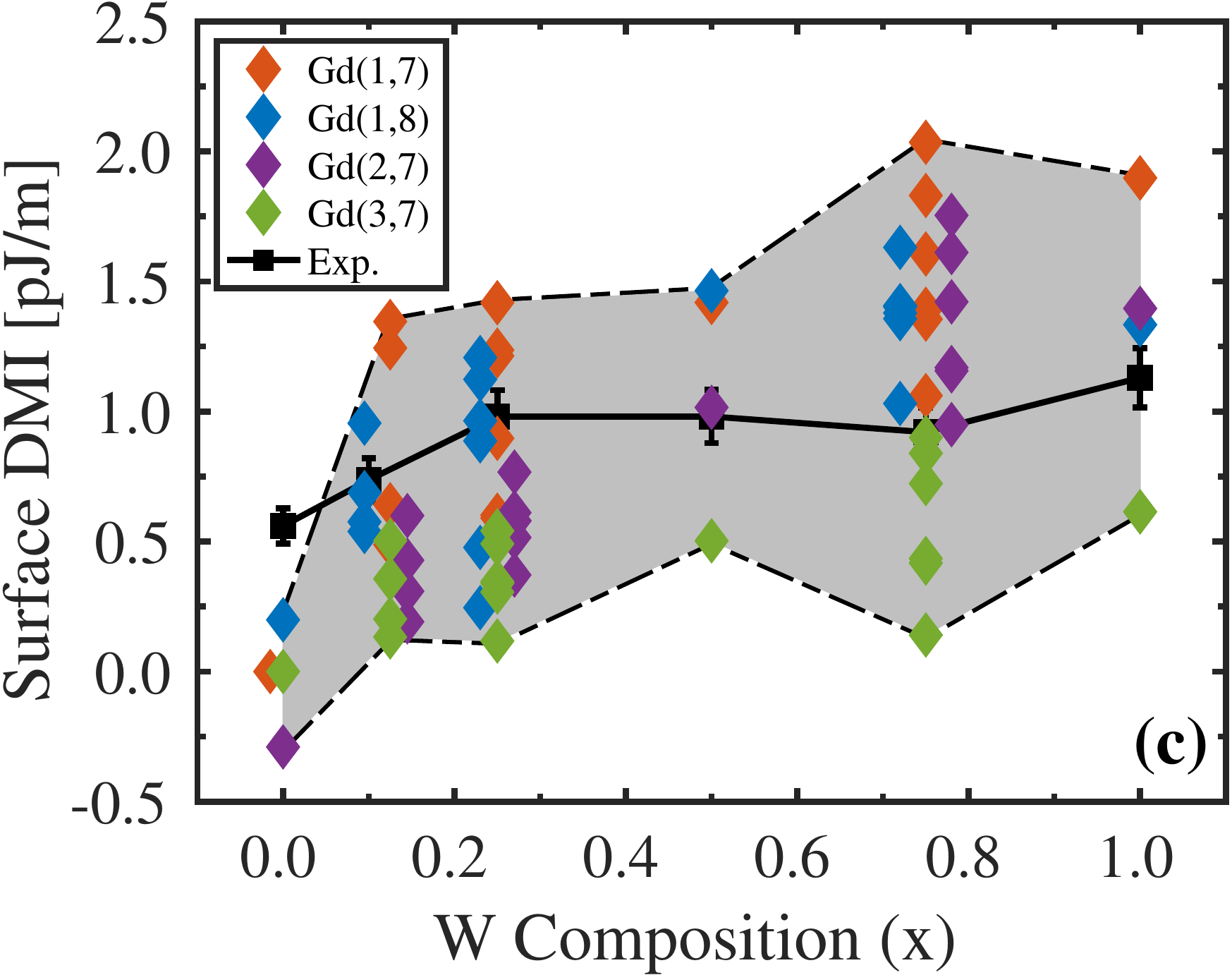}
\caption{The DMI as a function of W composition ($x$) in $\text{Pt}/\text{Gd}\text{Co}/\text{Pt}_{1-x}\text{W}_x$. (a) DMI variation with respect to W positions while the Gd atoms are fixed at $(1,7)$ positions. (b) Total spectrum of the DMI as both Gd and W positions are varied. For a specific W composition, each of the different colors represents the variation of Gd atomic positions, and the scattered points within the same color represent different W positions for that particular Gd arrangement in the structure (Fig.~\ref{structure}). (c) Surface DMI in comparison with experimentally observed DMI~\cite{yassine}. The numbers followed by the symbols Gd and W represent the positions of the respective atoms in the structure shown in Fig.~\ref{structure}.}
\label{iDMI}
\end{figure*}

\begin{figure}[!htbp]
\includegraphics[width=0.85\linewidth,scale=0.5]{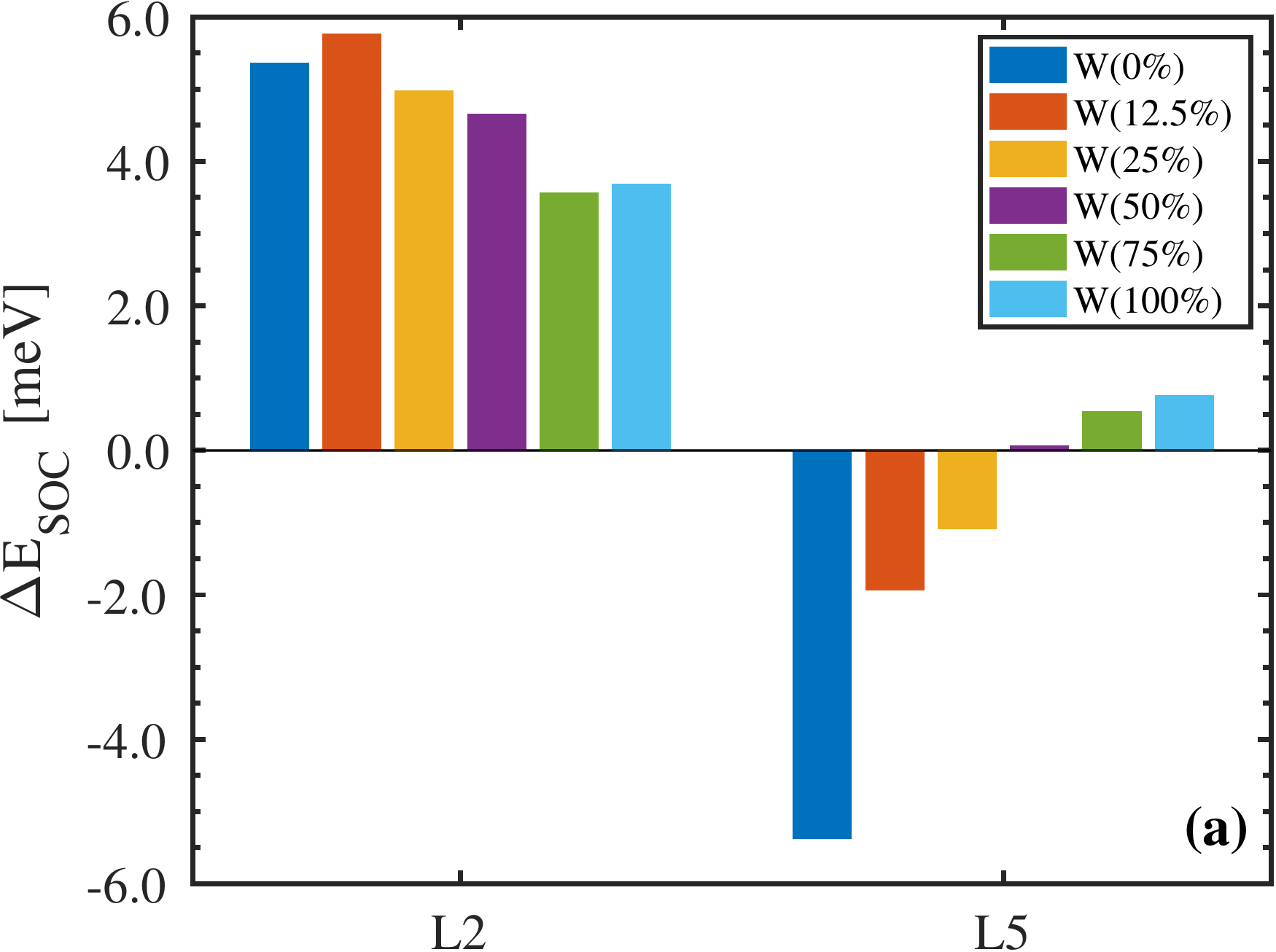}
\includegraphics[width=0.85\linewidth,scale=0.5]{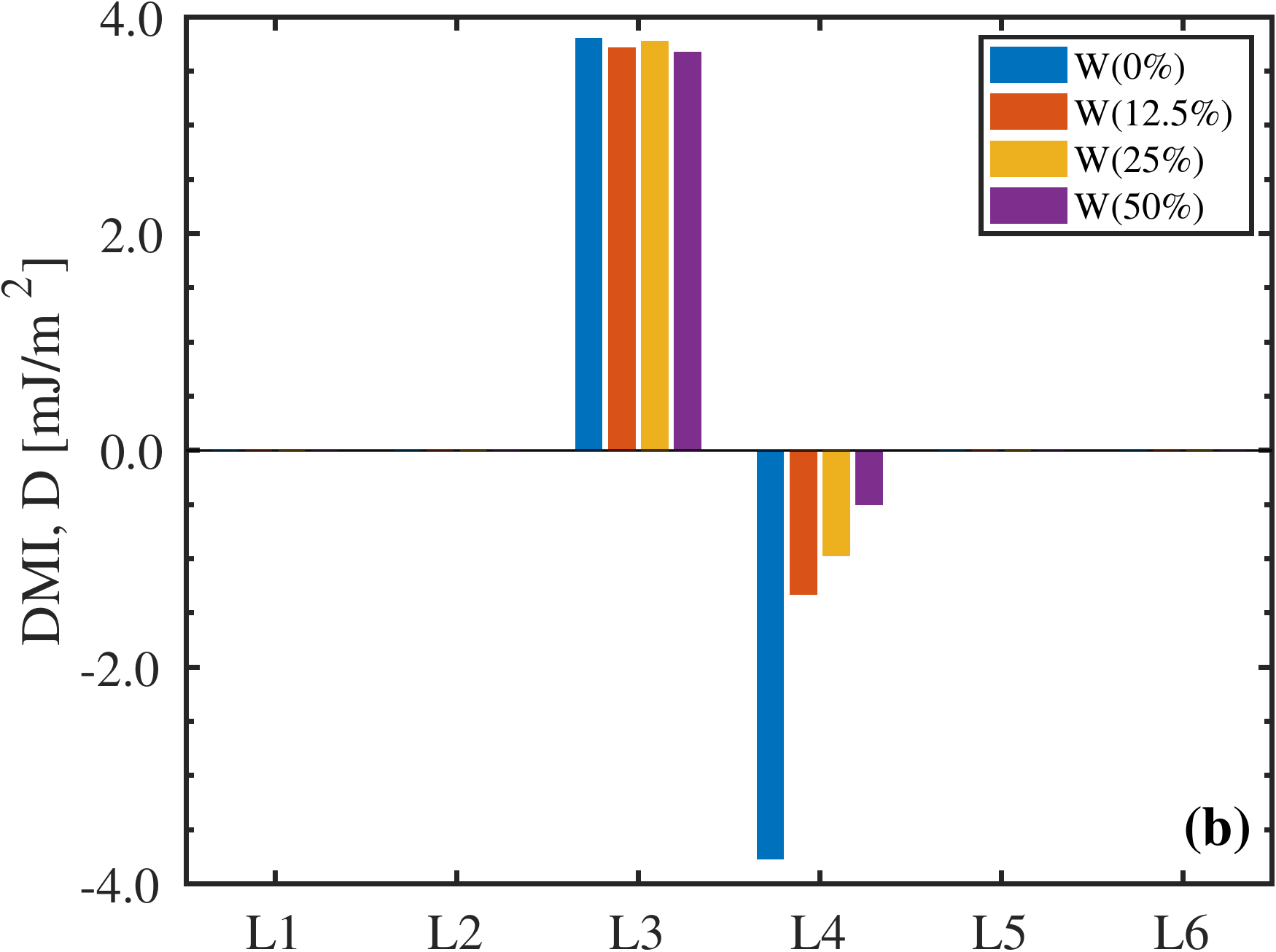}
\caption{(a) Change in SOC energy at the interfacial HM layers (L2 \& L5) as a result of changing spin chirality of the magnetic layers (L3 \& L4) from CW to ACW. All the color bars on the left (right) side represents the SOC energy change at L2 (L5) for different W compositions. (b) Layer resolved DMI for structures having W composition, $x=0\% - 50\%$.}
\label{SOC}
\end{figure}

\begin{figure}[!htbp]
\includegraphics[width=0.85\linewidth,scale=0.5]{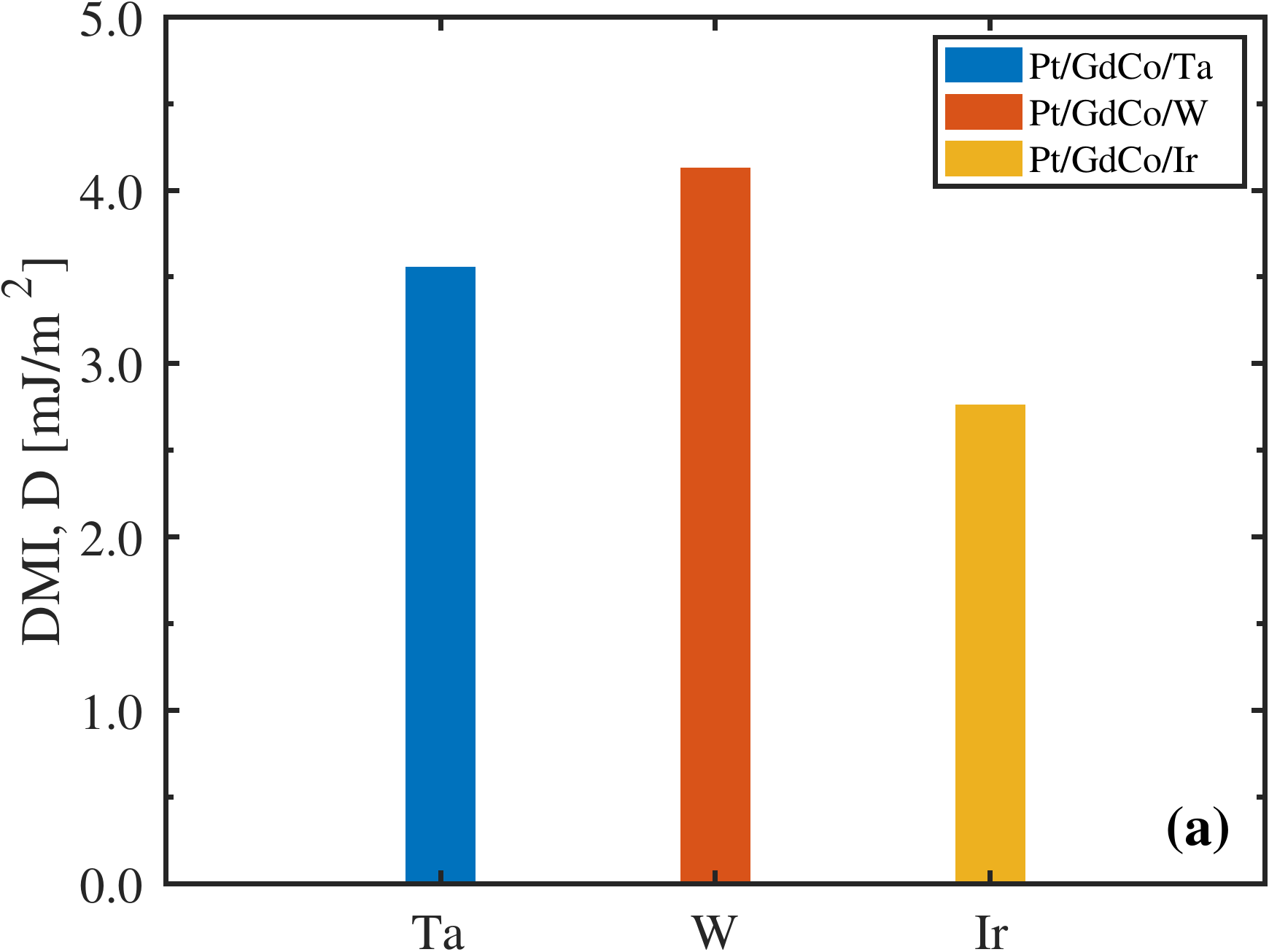}
\includegraphics[width=0.85\linewidth,scale=0.5]{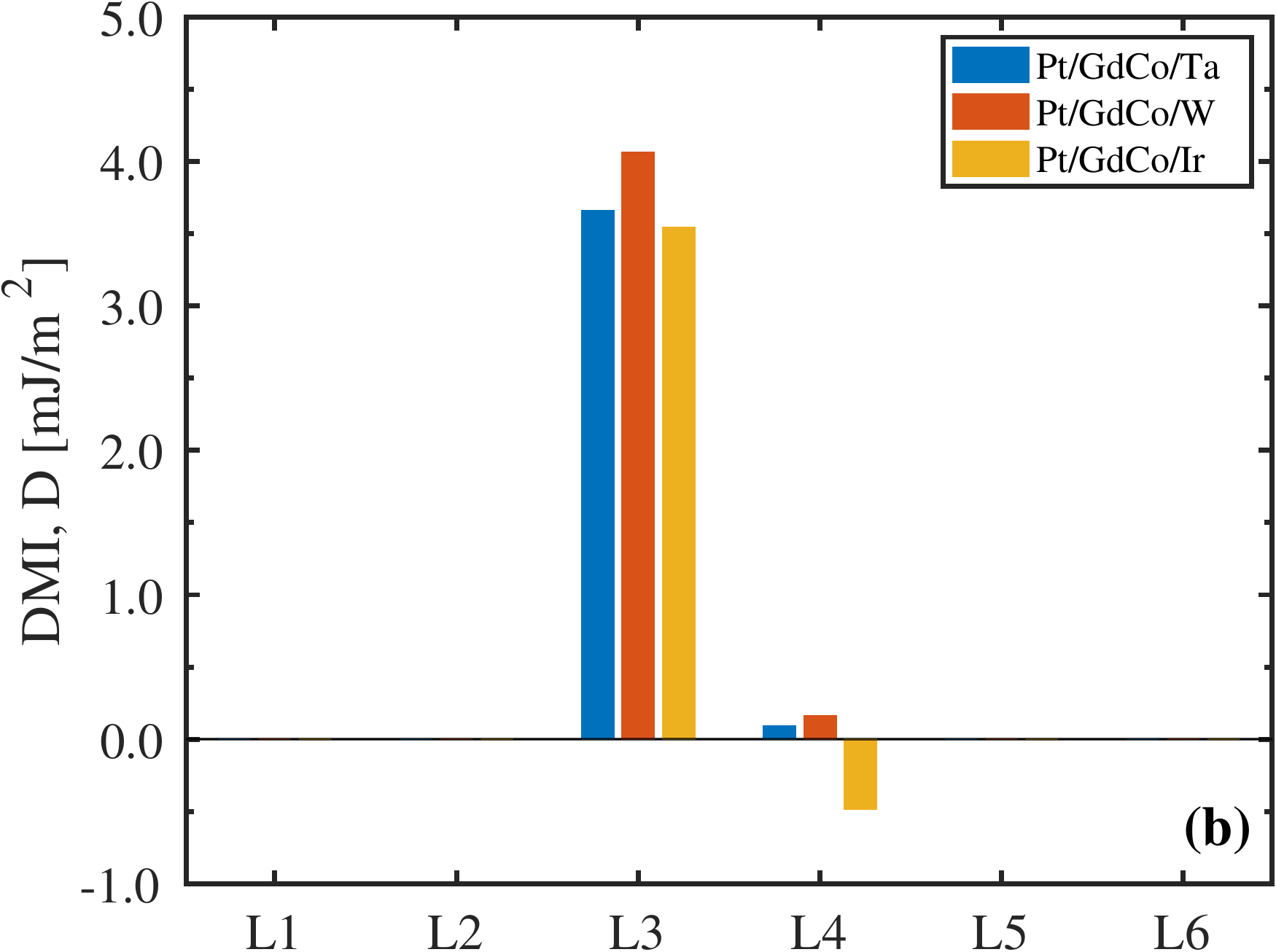}
\caption{(a) Calculated DMI in Pt/GdCo/X, where X=Ta, W, Ir. (b) Layer resolved DMI.}
\label{capping}
\end{figure}

\begin{figure}[!htbp]
\includegraphics[width=0.85\linewidth,scale=0.5]{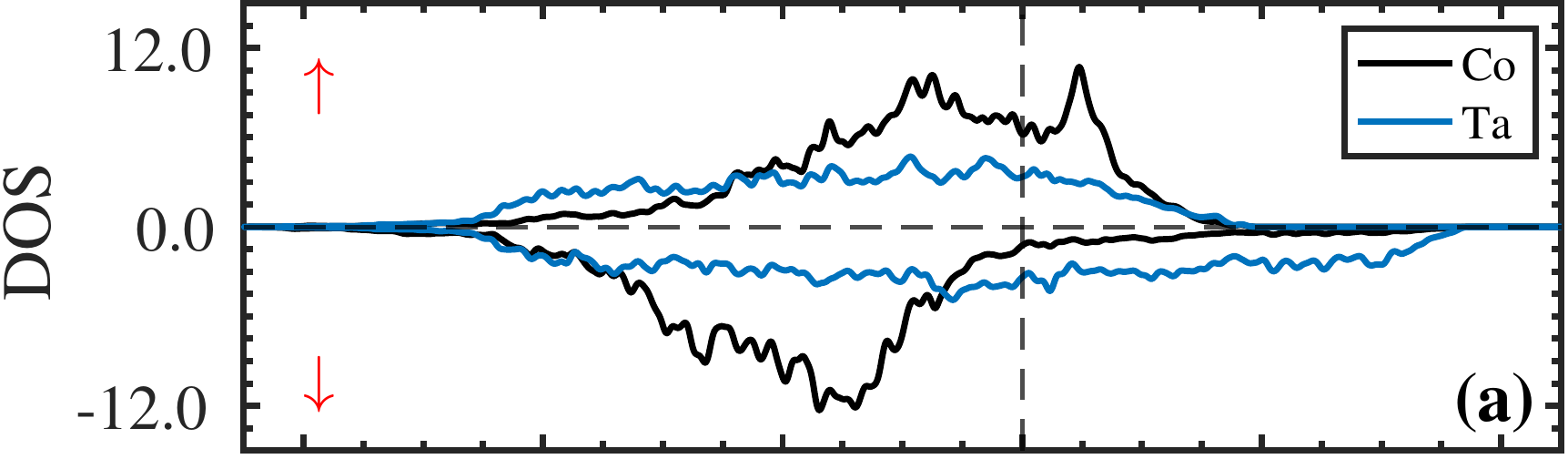}
\includegraphics[width=0.85\linewidth,scale=0.5]{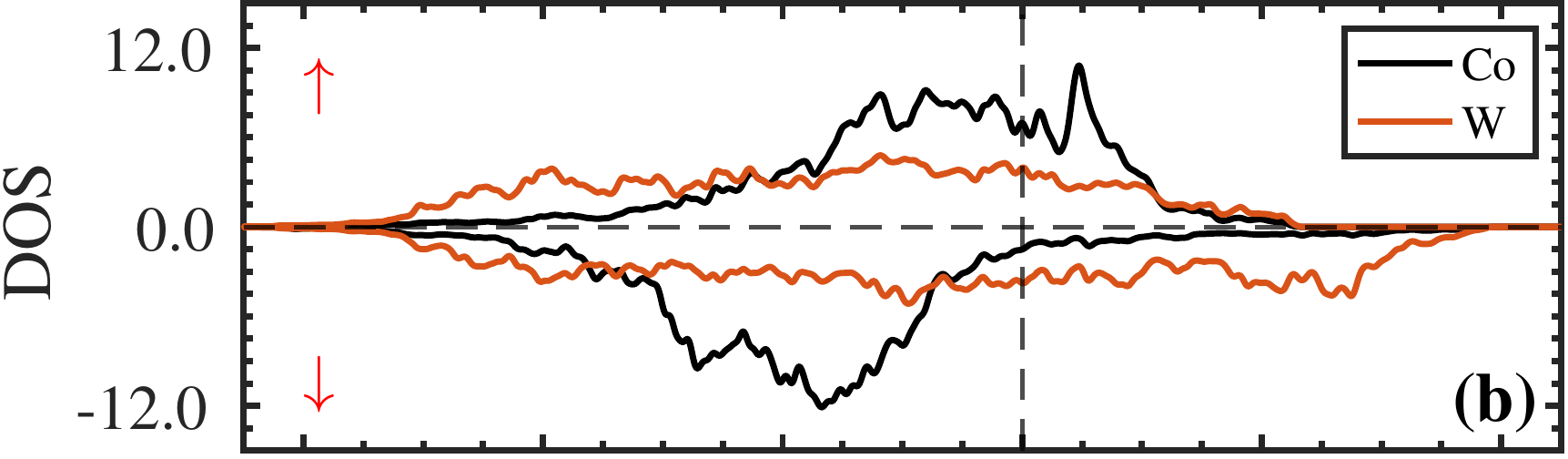}
\includegraphics[width=0.85\linewidth,scale=0.5]{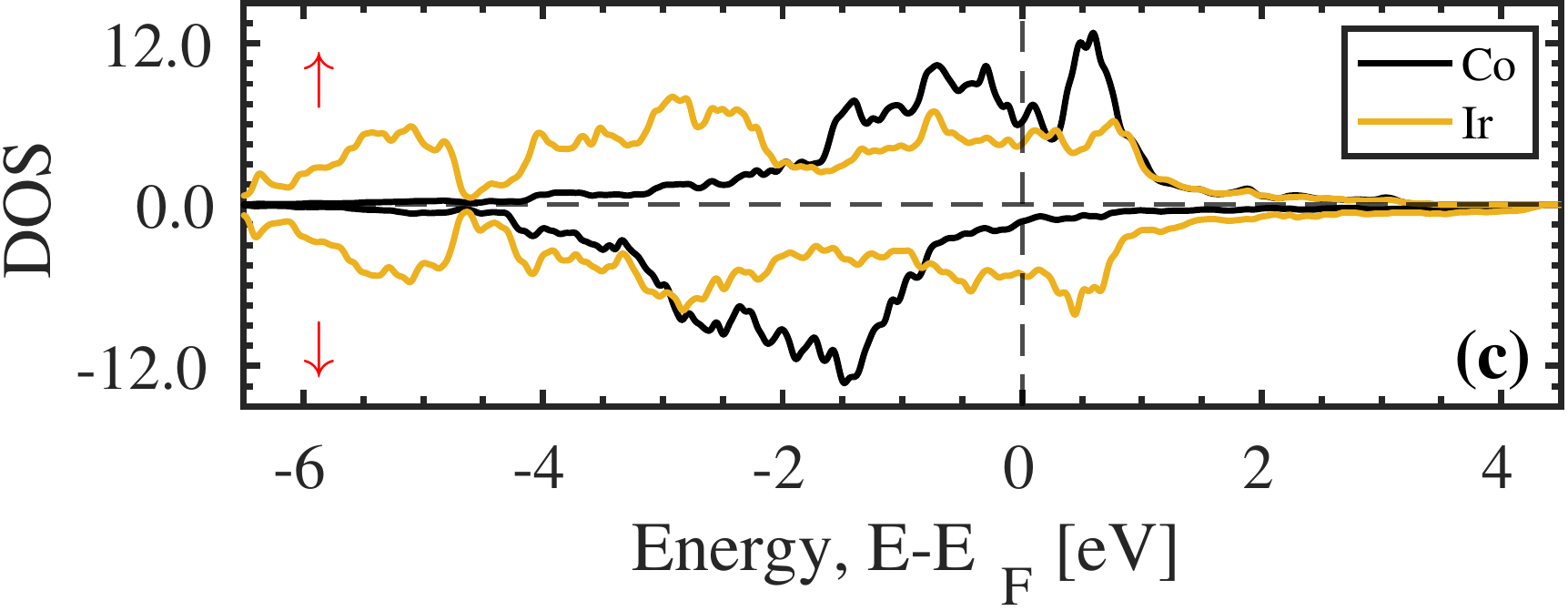}
\caption{Projected density of states (p-DOS) showing $3d$-$5d$ band alignment between Co (black) and X (colored) in Pt/GdCo/X. (a) X=Ta, (b) X=W, and (c) X= Ir. The red up (down) arrow represents the spin-up (spin-down) channel.}
\label{PDOS}
\end{figure}

 
\noindent {\it{Introduction.}} Magnetic skyrmions are topologically protected spin textures and are attractive for next-generation spintronic applications, such as racetrack memory and logic devices~\cite{Fert,app1,app2,app3,hamed1,hamed2,Sakib}. The interfacial Dzyaloshinskii-Moriya Interaction (DMI), an antisymmetric exchange originating from the strong spin-orbit coupling (SOC) in systems with broken inversion symmetry~\cite{DMI1,DMI2}, is one of the key ingredients in the formation of  skyrmions in magnetic multilayers~\cite{Heinze2011Jul,Moreau,amy}. Controlling the DMI offers the possibility to manipulate skyrmion properties, i.e., size and stability~\cite{size,stability}.\\
\indent Over the past few years, the underlying DMI physics and overall skyrmion dynamics have been studied extensively for ferromagnetic (FM) systems ~\cite{yang,Belabbes,Kashid,amy,Banerjee,Boulle,Tacchi}. Although both heavy metal (HM)/FM bilayers and HM/FM/HM sandwiched structures have been explored, most of the reported results are based on ideal interfaces. Indeed, very few studies focus on the role of disorder on DMI~\cite{Blugel}. Furthermore, ferrimagnetic materials have drawn attention due to their low saturation magnetization, low stray fields, reduced sensitivity to external magnetic fields, and fast spin dynamics, all of which favor ultra-fast and ultra-small skyrmions~\cite{Saima,Beach,Marco,Kim,Poon}. Very recently, Quessab \textit{et al.} have experimentally studied the interfacial DMI in amorphous Pt/GdCo thin films, and shown a strong tunability of the DMI by varying the thickness of the GdCo alloy and cap layer composition~\cite{yassine}. However, a detailed understanding of DMI, including the impact of two-sublattice ferrimagnetism, as well as the role of an experimentally realistic, chemically disordered interface are both missing.\\
\indent In this paper, we present a systematic theoretical analysis of the DMI in a compensated ferrimagnetic alloy using first principles calculations. In particular, we explore the variation of the DMI in $\text{Pt}/\text{Gd}\text{Co}/\text{Pt}_{1-x}\text{W}_x$ (Fig.~\ref{structure}) and find a strong tunability from $0$ to $4.42$ mJ/m$^2$ with variation in the W composition (Fig.~\ref{iDMI}). 
We studied the influence of atom placement and observed that the DMI is sensitive to structural variations such as the GdCo configuration in the thin magnetic film, and the PtW configuration at the interface. This is important to consider because, in reality, we have an amorphous alloy and the interfaces in deposited films are not perfect. We find a spectrum of DMI values that show an overall saturating trend, as seen in the experimental data \cite{yassine}. We argue that the change in SOC energy in the interfacial HM layers, especially the constancy of the SOC energy at the bottom layer and reduction of it in the cap layer, generates the observed saturating trend in the DMI with percentage of W incorporated (Fig.~\ref{SOC}). Additionally, we theoretically predict the variation of the DMI depending on the cap layer material, specifically for Pt/GdCo/X, where $\text X=\text {Ta}, \text W, \text {Ir}$ (Fig.~\ref{capping}). We find that the DMI is highest for $\text{W}$ in the cap layer and lowest for Ir, a trend that correlates with $3d$-$5d$ Co-X band alignment at the cap layer interface (Fig.~\ref{PDOS}). Our results identify the chemical and geometric factors responsible for interfacial DMI, and provide a potential path forward towards the engineering of material properties towards next generation skyrmion based spintronic applications.   

\noindent
{\it{Method.}} We use the technique of constraining the magnetic moments in a supercell to calculate the DMI within the Density Functional Theory (DFT) framework~\cite{yang}. The Vienna \textit{ab initio} simulation package (VASP) is used for the DFT calculations~\cite{vasp}. We use the projector augmented wave (PAW) potential to describe the core-electron interaction~\cite{PAW1, PAW2}. The Perdew-Burke-Ernzerhof (PBE) functional form of the generalized gradient approximation (GGA) is used for the exchange-correlation functional \cite{PBE}. In order to treat the on-site Coulomb interaction of Gd 4\textit{f}-electrons, we use the GGA+$U$ method~\cite{DFT_U} with an effective value of $U=6$ eV for Gd, as reported in previous studies for both bulk and slab calculations~\cite{Gd1, Gd2, Gd3}. We also validate the effective $U$ for our GdCo alloy by taking a range of $U$ values from $1-7$ eV, and confirming a stable ferrimagnetic ground state configuration of GdCo at $U=6$ eV. A $4\times1\times1$ supercell of $\text{Pt}(2)/\text{Gd}\text{Co}(2)/\text{Pt}_{1-x}\text{W}_x(2)$ (numbers in the parenthesis represent the number of monolayers) is used in all our calculations. While creating the $\text{Gd}\text{Co}$ alloy by replacing Gd atoms in the hcp Co(0001) slab, a $25\%$ Gd composition is maintained, which is the closest to the experimental proportion ($22\%$ Gd~\cite{yassine}) achievable within our structural arrangement. The trilayers are formed by aligning fcc(111) and hcp(0001) planes. The in-plane lattice constant of the slab structure is set to $2.81$ \si{\angstrom}, equal to the calculated nearest neighbor distance of bulk Pt, and the supercells are separated by a vacuum layer of $10$ \si{\angstrom} in the [001] direction. The cutoff energy is set to $500$ eV, and a $4\times16\times1$ Monkhorst-pack \textit{k}-grid is used for all the calculations. We verify the convergence of our calculations with cutoff energy, number of \textit{k}-points, and the thickness of the vacuum layer.\\
\indent The three step DMI calculation procedure starts with ionic relaxation along the atomic \textit{z}-coordinate to mimic a thin film, until the forces become less than $0.01$ eV/\si{\angstrom} and, the energy difference between two ionic relaxation steps becomes smaller than $10^{-6}$ eV. Next, in the absence of SOC, the non-spin polarized Kohn-Sham equations are solved to find an initial charge density. Finally, SOC is included, and the total energy of the system is calculated self-consistently for clockwise (CW) and anticlockwise (ACW) spin configurations (Fig.~\ref{structure}) until the energy difference between two consecutive steps becomes smaller than $10^{-6}$ eV.
\\
{\it{Results.}} The DMI energy ($E_{DMI}$) can be defined as
\begin{equation}
 E_{DMI} = \sum_{\langle i,j \rangle}{\textbf{d}_{ij} \cdot (\textbf{S}_i \times \textbf{S}_j)}
\end{equation}\\
    \noindent where $\textbf{S}_i$, $\textbf{S}_j$ are the nearest neighboring normalized atomic spins and $\textbf{d}_{ij}$ is the corresponding DMI vector. The total DMI strength, $d^{tot}$, defined by the summation of the DMI coefficient of each layer, to a first approximation, is calculated by the energy difference between the CW and ACW spin configurations~\cite{yang}, and expressed as $d^{tot}=(E_{CW}-E_{ACW})/12$. The micromagnetic DMI, D is given by $D=3\sqrt{2}d^{tot}/N_Fa^2$ ~\cite{yang}, where $N_F$ and $a$ represent the number of magnetic layers and the fcc lattice constant respectively.\\
\indent Before presenting the numerical results, it is worth mentioning that we can only investigate a limited subset of the structures for our calculations, as exploring all combinatorial possibilities is not feasible in terms of time and computational resources. We consider two separate alloy configurations: (i) Gd alloying in the magnetic layers, and (ii) W alloying in the cap layers.

In case (i), we first fix the position of the Gd atoms in the GdCo alloy. We maintain $25\%$ Gd composition separately in each magnetic layer, arguing that steric repulsion implies two Gd atoms are energetically unlikely to sit in the same layer, as assumed in previous studies~\cite{Nozaki}. The Gd atoms can thus arrange themselves in ${\binom{4}{1}}\times {\binom{4}{1}}=16$ ways. These sixteen combinations can be grouped into just four distinct sets because of their translational symmetry. In Fig.~\ref{structure}, looking at positions ($1-8$) in magnetic layers (L3 \& L4), it can be seen that Gd in $(1,7), (2,8), (3,5)$, and $(4,6)$ positions represent equivalent structures once the unit cell is periodically extended. Similarly, the other three groups are $[(1,8), (2,5), (3,6), (4,7)]$, $[(1,6), (2,7), (3,8), (4,5)]$, and $[(1,5), (2,6), (3,7), (4,8)]$. We confirmed this equivalence by calculating the energy of the Pt/GdCo stack by varying all the Gd positions and indeed find equal energy for the four structures within the same group. For case (ii), we choose one representative from each of the above four groups and proceed with W positional variations in the cap layer. While exploring W alloy configurations, for lower composition ($12.5\%-50\%$), W is only
incorporated in layer L6. Finally, we vary all the possible W positions and calculate the DMI for a total of $76$ structures.\\
\indent Figure~\ref{iDMI}(a) shows the calculated DMI, D for $\text{Pt}/\text{GdCo}/\text{Pt}_{1-x}\text{W}_x$, as a function of W composition. At $x=0\%$, the DMI vanishes as expected because, for a perfectly symmetric trilayer structure, the contributions from the bottom and top interfaces are equal and opposite. As the W composition increases from $0\%$ to $12.5\%$, we find a maximum DMI of $2.93$ mJ/m$^2$. The underlying mechanism behind this non-zero DMI is the inversion symmetry breaking of the $\text{Pt}/\text{GdCo}/\text{Pt}$ structure by the insertion of W atoms in the cap layer. We find that a small amount of W ($12.5\%$) gives a large DMI change, and subsequent to that initial rise, with increasing W content, the DMI saturates. As the composition of W increases, we find a maximum DMI of $4.42$ mJ/m$^2$ corresponding to $75\%$ W composition.\\
\indent We find that the DMI is very sensitive to the structural details, specifically the positions of the Gd and W atoms. Figure~\ref{iDMI}(a) shows the variation of D as the position of the W atoms changes. In Fig.~\ref{iDMI}(a), for all cases, Gd atoms are fixed at the $(1,7)$ positions. We show the variation of W positions for the structures with $12.5\%$, $25\%$, and $75\%$ compositions because for the other three cases there is only one combination possible in terms of W positions. Figure~\ref{iDMI}(b) shows the total spectrum of the DMI variation while varying both the Gd and W positions in the structure. Interestingly, for all the cases, the increasing trend of the DMI is very similar. We conjecture that changing the position of the atoms within the small unit cell will change the nature of the interface that gives variations in the DMI. For example, in the case of Pt/GdCo/W, when Gd atoms placed at position (1,7), the SOC energy change in the interfacial Pt layer is higher than that of position (3,7), which translates to the corresponding DMI as well.\\
\indent To validate our results against the recent experiment~\cite{yassine}, we calculate the surface DMI (in units of pJ/m) by multiplying the calculated DMI, D with the thickness of the magnetic layers. In our calculations, we use the thickness as $N_{F} a/\sqrt{3} = 4.6$ Å for the magnetic layers, while the experimental thickness is $5$ nm. Figure~\ref{iDMI}(c) shows the surface DMI from both the DFT calculation and the experiment, scaled by their respective thickness. In the experiment, a non-zero DMI of $0.56$ pJ/m (solid black line) is found for the Pt/GdCo/Pt structure because of the asymmetry in the bottom and the top interfaces due to the difference of interface roughness and intermixing \cite{yassine}. On the contrary in our DFT model, we use a perfect crystal structure that gives a near zero DMI for the symmetric cases (a small non-zero DMI might arise from intrinsic asymmetry within a thin crystalline GdCo film modeled here). We find an overall matching trend between the DFT and experimental data for the rest of the compositions. An exact quantitative agreement between the DFT results and the experiment is difficult to achieve because we use a crystal structure for our model, whereas, in the experiment, amorphous or polycrystalline materials are used. Additionally, the magnetization also differs between our model and the experiment as the thickness and the dimensions of the structure are different. However, we argue that the structural imperfections in the experiment amount to an ensemble averaging over the various configurations we theoretically explore, so that the experimental data falls in the middle of the spectrum (gray shaded area) of our DFT data.
\\
\indent In Fig.~\ref{iDMI}, the DMI increases non-monotonically as a function of W composition as opposed to a linear increase one may expect. This non-monotonic trend can be explained by the change of spin-orbit coupling energy, $\Delta E_{SOC}$, between CW and ACW spin configurations, in the HM layers adjacent to the magnetic layers in Fig.~\ref{structure}. In Fig.~\ref{SOC}(a), we show the $\Delta E_{SOC}$ in L$2$ (adjacent to the bottom magnetic layer) and L$5$ (adjacent to the top magnetic layer) for all W compositions ($0\%-100\%$). We find that $\Delta E_{SOC}$ in L$5$ changes drastically as W composition changes from $0\%$ to $12.5\%$, slowing down thereafter. On the other hand, distributions of $\Delta E_{SOC}$ in L$2$ are not very sensitive to the W composition. Although we find a relatively lower $\Delta E_{SOC}$ at L$2$ for $75\%$ and $100\%$ W compositions, the corresponding $\Delta E_{SOC}$s at L$5$ are positive. In trilayer structures, the DMIs of the bottom and top interface are additive~\cite{yang_sci, amy}, so that the sum arising from L$2$, and L$5$ accounts for the observed non-monotonic change of DMI in Fig.~\ref{iDMI}. From our findings, we conjecture that the inversion symmetry breaking plays a vital role on the DMI while the effect of W composition is not that prominent, in agreement with the recent experiment~\cite{yassine}.
\\
\indent To corroborate our analysis, we calculate the layer resolved DMI. Figure~\ref{SOC}(b) shows the layer resolved contribution of the DMI for the structures with $0\% - 50\%$ W composition. The results show that the DMI comes only from the interfacial magnetic layers. We can see that the change in the DMI contribution from the top interfacial layer (L$4$) with increasing W is small, generating a similar trend as $\Delta E_{SOC}$ shown in Fig.~\ref{SOC}(a). Additionally, the contribution from the bottom interfacial layer (L$3$) remains almost the same throughout the range of W compositions. The addition of the DMI from the bottom and the top interfaces produces a saturation in the overall DMI curve.
\\
\indent Finally, our theoretical model allows us to explore the tuning of DMI in ferrimagnetic systems with different cap layer compositions, which could be critical in designing suitable materials for hosting ultrasmall high-speed skyrmions. Furthermore, for applications, skyrmions can be driven by current-induced spin-orbit torques (SOT)~\cite{SOT}. Changing the cap layer HM offers the ability to tune the SOT efficiency and DMI simultaneously. We report the DMI of Pt/GdCo/X where $\text X=\text {Ta}, \text W, \text {Ir}$, to demonstrate the effect of cap layer 5\textit{d} transition HM on the DMI in Fig.~\ref{capping}(a). W and Ta are known for their giant spin-Hall angle~\cite{W_SHA, Ta_SHA}, and previous studies have shown an additive DMI for a ferromagnet sandwiched between Pt and Ir ~\cite{yang_sci,IrDMI}, which guides us to explore these structures and see which one of them has the largest DMI. We find that W in the cap layer favors higher DMI than Ta and Ir. To explain the DMI trend, we calculate the layer resolved DMI contribution from bottom and top interfaces, as shown in Fig.~\ref{capping}(b). From Fig.~\ref{capping}(b), we can observe that the DMI contribution from the top interface (L4) is large when Ir is used as a cap layer material while the DMI contributions are smaller for the cases of W and Ta. The observed trend of the DMI can be explained qualitatively by the Co 3\textit{d}-X 5\textit{d} band alignment, which controls the corresponding orbital hybridization. Figure~\ref{PDOS} shows the projected density of states (p-DOS) of Co-3\textit{d} and HM-5\textit{d} orbitals. Clearly, in Co/Ir, the band alignment around the Fermi level is higher than that of Co/W and Co/Ta, which in turn produce larger DMI contributions from L$4$ for Ir over W and Ta. The band alignment of Co/W and Co/Ta are close to each other. However, we note that the sign of the DMI contribution from the top interface is different for Ir than Ta and W. By analyzing the orbital projected densities of states of the cap layer HM, we find that Ta and W behave in a similar way i.e, $d_{xy}$ and $d_{x^2-y^2}$ have major contributions near the Fermi level while for Ir, the orbitals associated with the \textit{z} characters, namely $d_{xz}$, $d_{yz}$, and $d_{z^2}$ are prominent, correlating with the behavior shown in Fig.~\ref{capping}(b). Moreover, the variation of the DMI sign depending on the adjacent HM has previously been seen in both theoretical and experimental studies~\cite{Belabbes,Ma2018Apr}. Finally, adding the DMI contribution from both the interfaces (Fig.~\ref{capping}(b)) gives a smaller overall DMI for Pt/GdCo/Ir because of the large negative contribution from the top interface.\\
{\it{Conclusion.}} In summary, we demonstrate the impact of W composition in the cap layer of $\text{Pt}/\text{Gd}\text{Co}/\text{Pt}_{1-x}\text{W}_x$ trilayer structures using first principles calculations. We find excellent tunability of the DMI that shows a tendency of saturation with increasing W composition. The saturating trend of the DMI is attributed to the change of SOC energy at the top and the bottom intefacial HM layers as a function of W composition. Moreover, we find DMI sensitivity to the structural variation. We also demonstrate the DMI variation in Pt/GdCo/(Ta, W or Ir). We find W in the cap layer provides a higher DMI than Ta and Ir, due to the varying degree of orbital hybridization controlled by the band alignment between 3\textit{d}-5\textit{d} orbitals at the cap layer interface. Our results provide critical insights to the control mechanism of DMI in ferrimagnetic GdCo based systems, providing a path towards manipulating skyrmion properties for spintronic applications.\\
{\it{Acknowledgments.}} We thank Shruba Gangopadhyay, Jianhua Ma, Hamed Vakilitaleghani, and S. Joseph Poon for insightful discussions. This work is funded by the DARPA Topological Excitations in Electronics (TEE) program (grant D18AP00009). The calculations are done using the computational resources from High-Performance Computing systems at the University of Virginia (Rivanna) and XSEDE.
\bibliography{main,supporting}
\bibliographystyle{apsrev4-1}
\end{document}